# Increased peripheral lipid clearance in an animal model of Amyotrophic Lateral Sclerosis

Anissa FERGANI[1,2], Hugues OUDART[3], Jose-Luis GONZALEZ DE AGUILAR[1,2], Bastien FRICKER[1,2], Frédérique RENE[1,2], Jean-François HOCQUETTE[4], Vincent MEININGER[5], Luc DUPUIS[1,2]* & Jean-Philippe LOEFFLER[1,2]

1 :     Inserm, U692, Laboratoire de Signalisations Moléculaires et Neurodégénérescence, Strasbourg, F-67085 France
2 :     Université Louis Pasteur, Faculté de Médecine, UMRS692, Strasbourg, F-67085 France
3 :     Centre d'Ecologie et Physiologie Energétiques, UPR9010 CNRS, 23 rue Becquerel, 67087 Strasbourg Cedex, France
4 :     Equipe Croissance et Metabolismes du Muscle, Unite de Recherches sur les Herbivores, INRA, Centre de Clermont-Ferrand/Theix, 63122 St Genes-Champanelle, France
5 :     Fédération des Maladies du Système Nerveux, Centre référent maladie rare SLA, Hôpital de la Pitié-Salpêtrière, 47-83, Boulevard de l'Hôpital 75651 Paris, France

*       corresponding author

**Contact :**   Luc DUPUIS
                Phone: (+33) 3 90243091
                Fax:    (+33) 3 90243065
                e-mail: ldupuis@neurochem.u-strasbg.fr

**Running footline:**    lipid metabolism in amyotrophic lateral sclerosis mice

**Abstract :**

Amyotrophic lateral sclerosis (ALS) is the most common adult motor neuron disease causing motor neuron degeneration, muscle atrophy, paralysis and death. Despite this degenerative process, a stable hypermetabolic state has been observed in a large subset of patients. Mice expressing a mutant form of Cu/Zn superoxide dismutase (mSOD1 mice) constitute an animal model of ALS that, as patients, exhibits unexpectedly increased energy expenditure. Counterbalancing for this increase with a high fat diet extends lifespan and prevents motor neuron loss. Here we investigated whether lipid metabolism is defective in this animal model. Hepatic lipid metabolism was roughly normal while gastro-instestinal absorption of lipids as well as peripheral clearance of triglycerides-rich lipoproteins were markedly increased, leading to decreased postprandial lipidemia. This defect was corrected by the high fat regimen that typically induces neuroprotection in these animals. Altogether, our findings show that energy metabolism in mSOD1 mice shifts towards an increase in the peripheral use of lipids. This metabolic shift probably accounts for the protective effect of dietary lipids in this model.

**Supplementary keywords:**

Plasma lipoproteins, neurodegeneration, motor neuron, LDL, HDL, liver metabolism, intestinal absorption, skeletal muscle

**Introduction**

Neurodegenerative diseases have been long considered as the result of locally restricted injury to specific neurons, the loss of which represents both the origin and the end of the pathological process. However, this "cell-autonomous injury" hypothesis does not seem to hold true for disorders in which neurodegeneration would be rather caused by the combined action of a series of defects arising at the level of the whole organism and eventually leading to a very selective cell death. Although still poorly understood, peripheral metabolic abnormalities, as shown to occur in several neurodegenerative diseases (1-5), could participate in the neurodegenerative process (6, 7).

Amyotrophic lateral sclerosis (ALS) is a neurodegenerative disease characterized by the progressive loss of motor neurons in the spinal cord, brainstem and motor cortex. Although tremendous efforts have been devoted to explain ALS pathogenesis, why motor neurons selectively die in this condition is still unsolved (8, 9). Animal models of ALS displaying the major features of the human disease are transgenic mice overexpressing mutant forms of Cu/Zn-superoxide dismutase (SOD1) (10-12), a free radical scavenging enzyme that protects cells against oxidative stress and is mutated in a subset of patients with autosomal dominantly inherited ALS (13). Studies using these mice have postulated that mutant SOD1 (mSOD1) triggers ALS by a non-cell-autonomous mechanism involving not only motor neurons themselves but also other environing cells such as astrocytes and microglia (14, 15).

Beyond the neuromuscular system, we recently observed systemic abnormalities occurring in animal models of ALS (6, 16). mSOD1 mice are leaner than wild-type littermates and their fat pads gradually deplete as a result of a prominent hypermetabolic trait mainly of muscular origin (6). In line with these findings, higher resting energy expenditure

Fergani et al. 3/37/

has also been observed in a large subset of ALS patients (17, 18). Moreover, a significant percentage of patients present with glucose intolerance, and increased rates of muscle glucose uptake, oxygen consumption and lactate output that indicate the presence of marked abnormalities in carbohydrate metabolism in muscle tissue (19). These metabolic alterations that, at least in mice, precede motor neuron death, are not only associated but also contribute to the neurodegenerative process, since it has been shown in mSOD1 mice that increasing the energy content of the diet prolongs lifespan and maintains motor neuron numbers (6, 20) while restricting calorie intake hastens the disease (21).

The origin of the systemic defects in mSOD1 mice and their contribution to the pathology in ALS patients needs to be elucidated. Given that a highly energetic fat regimen protects against the disease in animal models, we hypothesize that compensation of as yet unidentified disturbances in lipid metabolism could account for the benefits of such a regimen. In the search for these lipid disturbances, we show here that mSOD1 mice present with decreased post-prandial lipidemia characterized by increased peripheral clearance of triglycerides-rich lipoproteins, probably caused by skeletal muscle hypermetabolism. This hypolipidemia was reverted by the neuroprotective high fat regimen. Altogether, our findings show that energy metabolism in mSOD1 mice shifts towards an increase in the peripheral use of lipids. This metabolic shift probably accounts for the protective effect of dietary lipids in this model.

## Materials and Methods

### Animals

Transgenic female mice with the G86R murine SOD1 mutation were maintained in a *FVB* background and maintained with their nontransgenic age-matched female littermates on a 12 hr light/dark cycle. They were fed with a chow diet (A04, UAR, Epinay sur Orge, France) unless otherwise stated and had free access to water. High fat diet consisted of A04 diet complemented with 20% butter fat. Transgenic mice expressing human G93A SOD1 mutations were kindly provided by Faust pharmaceuticals.

### Lipid measurements and Lipoprotein fractionation

Tail vein blood was collected in heparinized capillary tubes, placed on ice and centrifuged at 3000g, 4°C. Plasma levels of triglycerides (TG), cholesterol and total lipids were determined with Randox (Crumlin, UK) kits TR213, CH200 and TL100 respectively. The TG TR213 kit is an enzymatic kit based upon the hydrolysis of TGs in glycerol and fatty acids, the glycerol produced being oxidized to reach $H_2O_2$, which oxidizes 4-aminophenazone and 4-chlorophenol to reach a colored quinoneimine compound. The cholesterol CH200 kit is based upon the oxydation of cholesterol to cholestene 3-one and $H_2O_2$ by the cholesterol oxidase. The $H_2O_2$ produced is then dosed upon the same principle than in TR213 kit. In the CH kit, the enzymatic reagent also includes cholesterol esterase, thus allowing the dosage of not only free but also esterified cholesterol. Both TR213 and CH200 kits yielded results in our hands similar to those obtained in a certified clinical biochemistry laboratory (data not shown). The TL100 kit is based upon the reaction of lipids with sulphuric acid, phosphoric acid and vanillin to form a pink colored complex. This kit was able to accurately detect exogenous triglycerides added in serum (data not shown). Moreover, this kit yielded results consistent with those obtained using gravimetric detection of lipids following Folch extraction (data not shown). In all cases, a 5-point standard curve was performed using the standards provided in the kits and results were analyzed only if the correlation coefficient of the standard curve was above 0,995. Serum concentrations were calculated using the equation of the regression line as calculated by Excel (Microsoft). In our hands, intradosage variability of these three kits was below 5%. Serum lipoprotein profiling was performed by the clinical mouse institute, (Strasbourg, France) by fast performance liquid chromatography on Dionex (Sunnyvale, CA, USA) apparatus.

**Lipid and glucose intestinal absorption**

Lipid and glucose uptake rate into the small intestine was determined by using the *in vivo* perfused intestinal segments technique (22). Normally fed mice were anaesthetized and placed on a heated (37°C) surgical table. After performing a laparotomy, a 10 cm segment of the small intestine was canulated and the luminal contents removed by gently flushing with saline solution at 37°C. An intestinal loop was cannulated and a recirculating perfusion was started at a flow rate of 2ml $min^{-1}$ with Intralipid solution (4%) (Pharmacia, Strasbourg, France) at 37°C for triglycerides uptake. TG and NEFAs concentrations were estimated in the luminal content using Randox kits TR213 and FA115 respectively. For glucose uptake, a saline solution at 37°C containing 5mM glucose was perfused. The absorption rate was calculated as the difference between intial and final glucose concentrations using a glucometer (Accu check, Roche, Basel, Switzerland) following manufacturer's instructions.

**Gastric emptying and intestinal transit**

Gastric emptying time and intestinal transit were determined using methylene blue as a tracer dye. Briefly, after a four-hour fast, mice were gavaged with 0.1mL of methylene blue/10% dextrose solution and sacrificed 30 minutes after gavage. The stomach was clamped above the oesophageal sphincter and below the pylorus to prevent the leakage of the dye. Stomachs were cut and immediately homogenized in 10 mL of NaOH 0.1 M. After clearing steps by centrifugation, OD at 562nm of the supernatant was determined. Gastric emptying was determined as the difference between the measured OD and those of a group of mice, gavaged in parallel and sacrificed 1 min after gavage. Intestinal transit was determined as the most distal point of migration of methylene blue in the intestine and expressed as a percentage of the total length of the intestine.

**RT-PCR analysis**

Each frozen sample (livers or intestinal mucosa) was placed into a tube containing a 5-mm stainless steel bead (Qiagen, Courtaboeuf, France) and 1ml of Trizol reagent (Invitrogen, Paisley, UK) and homogenised using a TissueLyser (Qiagen, Courtaboeuf, France) at 30 Hz for 3 min. RNA was extracted by the chloroform/isopropyl alcohol/ethanol (Sigma, Lyon, France) method and stored at −80°C until use. RNA reverse transcription and Sybrgreen realtime PCR assays were performed using the Biorad (Marnes la Coquette, France) iCycler, kits and protocols. Primer sequences are available in supplementary table 1. The relative levels of each RNA were normalized to the corresponding 18S RNA levels.

**Hepatic VLDL production**

After a 4 hour fast, mice were injected intravenously into the tail vein with 20 mg of Tyloxapol (Sigma, Lyon, France) as a 20% solution in PBS buffer. Blood samples were drawn at 0, 60, 120, and 240 min after the Tyloxapol injection, and TG concentrations were determined in plasma as described above.

**Intragastric fat load**

After a 4 hour fast, mice were given intragastrically a 400 µl olive oil bolus. Blood samples were drawn at 0, 1, 2, 3, 4, and 5 h after bolus administration, and TG concentrations were determined in plasma as described above.

**Histological analysis**

Adult mice were intracardiacally perfused with paraformaldehyde 4% (Sigma, Lyon, France). Liver were immediately post-fixed in paraformaldehyde 4% during 24h. Serial sections of 40 µm were obtained by vibratome. To visualize fat vesicles, slides were incubated in Oil Red O solution (0.1% Oil Red O dissolved in isopropanol, Sigma, Lyon, France)) for 15 min at room temperature, washed with 30% isopropanol (Sigma, Lyon, France) and distilled water.

**Lipoprotein lipase activity**

Lipoprotein lipase (LPL) activity was assessed after homogenization of the tissues in a buffer composed of ammonia-HCl (25 mM) pH 8.2, containing EDTA (5 mM), Triton-X-100 (8 g/l), sodium dodecyl sulfate (0.10 g/l), heparin (Sigma, Lyon, France) 5000 IU/l) and peptidase inhibitors(Sigma, Lyon, France). Insoluble material was discarded by centrifugation at 20000g for 20 min at 4°C. As described by Hocquette and collaborators (23), rat serum was used as activator, and Intralipid® (Pharmacia, Strasbourg, France), into which [3H] triolein (Perkin Elmer, Jügesheim, Germany) has been incorporated, was used as the substrate. Liberated [3H]-free fatty acids were quantified by liquid scintillation.

**Statistical analysis**

Statistical comparisons of these data were performed either using unpaired Student's *t* test or One-way anova followed by post-hoc Newman-Keul's test, using the GraphPad Prism software (version 4). Significance was considered below p=0.05.

# Results

**mSOD1 mice display alterations in blood lipid levels**

We had previously shown that mSOD1 mice lose progressively their adipose tissue stores and that feeding them with a diet enriched in fat prolonged lifespan and rescued motor neurons (6). Here we observed that mSOD1 mice displayed normal levels of plasma triglycerides (TG) and cholesterol under fasting but decreased levels of plasma TG, cholesterol and total lipids under normal feeding (figure 1A-C). To ascertain whether the observed decrease in lipid levels in mice was due to decreased postprandial lipidemia, we gavaged mice with olive oil and measured TG levels 3 hours later. In this experimental setting, we observed a roughly 2-fold decrease in TG levels as compared to wild-type littermates (figure 1A, right panel), thus showing that mSOD1 mice, although normolipidemic, present with decreased postprandial lipidemia. In order to assess whether this defect could be corrected by a high-fat diet (HFD), we fed a group of mice with 20% butter fat during four weeks and evaluated their lipid metabolism. As previously described (6), HFD reverted the deficit in body mass and repleted adipose tissue stores in mSOD1 mice. In addition, consistent with the documented protective role of this regimen, after one month of HFD feeding, only 1 of 10 mice displayed motor troubles while 5 out of the 9 mice normally fed showed ALS symptoms (data not shown). The high fat regimen also abolished the decrease in postprandial cholesterolemia (figure 1D). Altogether, these findings indicate that correcting the decreased postprandial cholesterolemia by feeding mSOD1 mice with a high fat regimen is associated with attenuation of ALS symptoms.

**mSOD1 mice have a defect in triglycerides-rich lipoproteins**

The main difference between fasting and feeding in terms of TG is the presence of chylomicrons in the blood of fed animals. Therefore, the decrease in TG in fed but not fasted mSOD1 mice suggested that the chylomicron system of lipoproteins was affected by the presence of mSOD1. To determine whether other lipoprotein fractions were affected, we performed fast performance liquid chromatography (FPLC) to separate the different lipoproteins and quantified their cholesterol content. We noted a strong decrease (roughly 50%) in VLDL and LDL cholesterol fractions in 75-days old asymptomatic mSOD1 mice, but only a slight decrease (less than 20%) in HDL cholesterol (figure 2A-B). Thus, the activity of the two lipoprotein systems involved in the transport of lipids towards peripheral tissues, namely chylomicrons, carrying dietary fats, and VLDL/LDL lipoproteins, transporting endogenously synthesized lipids, appears decreased in fed asymptomatic mSOD1 mice. The reverse transport of lipids by HDL lipoproteins is only relatively modestly affected. Interestingly, FPLC profiles of wild-type and mSOD1 mice were indistinguishable after 4 hours of fasting (figure 2C), which suggests that normally fed mSOD1 mice behave as they were abstained from nutriments, even though they ingest equal or higher amounts of food than their wild-type littermates (6).

**Intestinal absorption of lipids is increased in mSOD1 mice**

The observed decrease in circulating lipids might be caused by different mechanisms. First, mSOD1 mice might have a decreased food intake. We have already excluded this possibility in asymptomatic mice (6). Second, it is possible that the gastrointestinal system is malfunctioning. This hypothesis is further substantiated by clinical reports of gastrointestinal dysfunction in ALS patients (24). Indeed, mutant SOD1 was highly expressed in intestinal mucosa with no age-dependent variations that could explain gastrointestinal dysfunction (figure 3A). However, gastric emptying and intestinal transit after an oral gavage of methylene blue under fasting conditions appeared unchanged between mSOD1 mice and wild-type littermates (figure 3B-C), thus excluding gross gastrointestinal dysfunction. Third, nutrients could be poorly absorbed in the gut. To directly evaluate the absorptive potential of mSOD1 intestine, we used the *in vivo* perfused intestinal segment technique (figure 3D) and perfused either diluted intralipids, a TG-rich solution (figure 3E-F) or glucose (figure 3G). Glucose uptake by the intestine was identical between SOD1 and wild-type mice, and served as control. In contrast, levels of TG in the perfusate after intralipid infusion were constantly lower in mSOD1 intestines than in wild-type littermates analyzed in parallel (figure 3E). TG are broken down in the intestine to yield non-esterified fatty acids (NEFA), which are the molecular species absorbed by enterocytes (figure 3D). Our data then suggested that TG were degraded more rapidly in mSOD1 mice than in wild-type littermates. To determine whether NEFA generated from TG were efficiently absorbed and not accumulated in the gut, we measured them in intestines perfused with a TG-rich solution. In these experiments, the steady state levels of NEFA in intestinal perfusates were constantly lower in mSOD1 mice as compared to wild-type animals (figure 3F), thus showing that the process of NEFA absorption was not impaired but rather increased in mSOD1 mice. Together with the increased food intake displayed by mSOD1 mice (6) and the absence of gross dysfunction in gastric emptying and intestinal transit, the present data rule out the possibility that decreased lipidemia in fed animals was due to diminished intestinal absorption.

**Liver lipid metabolism is normal in mSOD1 mice**

A second mechanism that could account for decreased lipidemia in mSOD1 mice could be a shift in hepatic lipid metabolism, triggered by the high expression of mSOD1 in liver (figure 4A). Less release of VLDL from liver, caused by decreased TG or cholesterol biosynthesis or by impaired release itself, could in fact explain the decrease of VLDL fractions observed in mSOD1 mice. We measured the mRNA levels of fatty acid synthase (FAS) and Serum responsive element binding protein 1(SREBP1), involved in hepatic TG biosynthesis (25), and found that they were unchanged in presymptomatic mice, when hypolipidemia was already detectable (figure 4A). Contrastingly, expression of SREBP1 (and that of FAS at a lesser extent) was downregulated in the liver of symptomatic mice, which is consistent with the documented decrease in insulin levels and the general metabolic shutdown observed in mSOD1 mice at disease onset (6). As far as cholesterol metabolism is concerned, we did not detect any change in the expression of key genes that could compromise cholesterol biosynthesis. Indeed, expression of HMG-CoA reductase, the rate limiting enzyme in cholesterol biosynthesis, was rather increased in presymptomatic mice, whereas mRNA levels of SREBP2, a key transcription factor in this pathway, were unchanged before motor symptoms appeared and increased in end-stage mice (figure 4A). Altogether, these data thus suggest that decreased post-prandial lipidemia in mSOD1 mice cannot be ascribed to a deficiency in the expression of key enzymes or transcription factors controlling lipid biosynthesis.

An alternative mechanism leading to decreased post-prandial lipidemia could be an increase in lipid catabolism in liver that would generate elevated concentrations of ketone bodies. This does not seem to be the case since mRNA levels of carnitine palmitoyl transferase IA (CPT1A), the rate limiting enzyme in fatty acid oxidation, were unchanged in asymptomatic animals (figure 4A). In contrast, the increased expression of CPT1A in diseased mice is consistent with the high amounts of ketone bodies found in plasma at this stage (6). Another explanation might be that cholesterol is excreted as bile acid thus leading to hypolipidemia. However, the expression of cholesterol 7 alpha hydroxylase (CYP7A1), which is correlated to bile acid biosynthesis (26), was unchanged in presymptomatic mice (figure 4A), making therefore unlikely that an increase in bile acid excretion could be the cause of the early installed decreased post prandial lipidemia. It should be also noted that treating mSOD1 mice with cholestyramine, a drug known to stimulate bile acid secretion, was able to trigger an increase in CYP7A1 expression, thus showing that transcription could be still induced in these mice in response to an appropriate stimulus (data not shown). Finally, the expression of genes involved in lipoprotein assembly, such as ApoE or the transcriptional coactivator PGC1-β, appeared unchanged in presymptomatic mice and increased in diseased animals (figure 4A), which further reinforces the notion that lipoprotein assembly is not deficient at the transcriptional level in mSOD1 mice. Interestingly, mRNA levels of the LDL receptor (LDLR) were increased at presymptomatic stage, suggesting that the turnover of LDL might be increased in mSOD1 mice.

Hepatic VLDL assembly and secretion could be also impaired post-transcriptionnally thus leading to hypolipidemia, as occur, for instance, when nascent VLDL are retained in liver. To test whether fats accumulate in the liver of mSOD1 mice, we visualized lipidic droplets by Oil Red O staining. As illustrated in figure 4B, accumulations of lipids were clearly distinguishable and comparable between wild-type and mSOD1 mice, although an overall fainter stain could be observed in end-stage mice. In all, indirect evidences all pointed out to a normal hepatic lipid metabolism in mSOD1 mice

**TG clearance by peripheral tissues is increased in mSOD1 mice**

To directly test whether mSOD1 mouse liver releases VLDL efficiently, we measured hepatic VLDL production under fasting conditions using tyloxapol (Triton WR 1339), a detergent that coats lipoprotein complexes and thus impairs their peripheral clearance. In tyloxapol-injected mice, the rate of increase in plasma TG is therefore correlated to the biosynthesis of VLDL in liver. The increasing concentrations of TG in plasma were almost identical in mSOD1 and wild-type mice (figure 5A), thus showing that VLDL production was not impaired in the transgenic animals. Altogether, these data show that decreased post-prandial lipidemia in presymptomatic mSOD1 mice cannot be explained by the altered capacity of liver to synthesize TG and release them in the form of VLDL into the circulation.

To examine whether the drop in plasma TG-rich lipoproteins was due to increased peripheral uptake of lipids, we performed a fat-loading test by administrating TG intragastrically in the form of olive oil after a 4-hour fast. Initial TG levels were similar in mSOD1 and wild-type mice, consistent with previous results (figure 5B). TG levels increased gradually and comparably between the two groups during the first two hours after gavage, which further confirms that intestinal absorption of fat as well as chylomicron production by the intestine are roughly normal in the transgenic animals (figure 3). Later, the clearance of TG, represented by the decreasing segments of the curves in figure 5B, was faster in mSOD1 mice than in wild-type littermates. These data reflect the increased ability of mSOD1 mice to metabolize TG. We propose that this is likely to be the cause of the decreased post prandial lipidemia observed in these animals.

**Gene expression changes point to muscle as the origin of increased lipid peripheral uptake**

Since we had previously shown that skeletal muscle is characterized by an hypermetabolic trait in mSOD1 mice (6), and is one of the most important tissues that captures lipids, we tested whether muscle could be the site of increased lipid uptake. Surprisingly, the expression levels of lipoprotein lipase (LPL) were unchanged in asymptomatic mSOD1 mice while increased in symptomatic animals (figure 6A, left panel). This late increase does not seem to be the result of denervation, since LPL mRNA levels did not increase but rather decrease following sciatic nerve axotomy (figure 6A, right panel). In order to test whether muscle lipid uptake could be promoted by increased LPL activity, we measured total LPL enzymatic activity but did not find any difference between mSOD1 and wild-type mice (figure 6B). However, mRNA levels of other several genes involved in lipoprotein clearance, such as LDL receptor, VLDL receptor and the fatty acid transporter FAT/CD36, were augmented in the muscle of both presymptomatic and diseased mice (figure 6C), which supports the hypothesis that muscle tissue is a primary site of increased lipid consumption in mSOD1 mice. To exclude the possibility that our observations could result from a transgenic artifact, we also measured the expression of LPL, vLDLR, LDLR and FAT/CD36 in muscles of G93A mice, another transgenic line overexpressing a human mutant SOD1 (11). We found in these mice not only the previously observed upregulation of vLDLR and FAT/CD36 but also increased levels of LPL mRNA in presymptomatic mice (figure 6D). Altogether, these data suggest that decreased post prandial lipidemia in mSOD1 mice is driven by the increased clearance of TG in peripheral tissues, particularly skeletal muscle.

# Discussion

We show here that energy metabolism of mSOD1 mice, an animal model of ALS, is shifted towards an increased peripheral use of lipids. This metabolic shift probably accounts for the protective effect of dietary lipids in this model.

We had previously shown that mSOD1 mice are afflicted by prominent unbalanced energy homeostasis. In particular, two different mSOD1 mouse lines display increased resting energy expenditure as determined by indirect calorimetry (6). Importantly, 30 to 60% of ALS patients also present with hypermetabolism as determined by the same methodology (17, 18, 27). We then suggested that the origin of this hypermetabolic trait was an increased demand for nutrients in muscle tissue. This hypothesis was supported by increased glucose uptake and altered gene expression of enzymes involved in glucose and lipid use that we observed in muscles of presymptomatic mSOD1 mice (6). Our present data now show that the lipids supplied by normal feeding rapidly disappear from plasma, pointing to muscle hypermetabolism as triggering increased peripheral clearance of TG-rich lipoproteins. This phenomenon, associated with unchanged levels of VLDL secretion by the liver, is likely to account for the postprandial hypolipidemia observed in the transgenic mice.

The results reported herein, along with our previous studies (6, 16), support the dysfunction of skeletal muscle metabolism as the cause of the impairment of energy homeostasis in mSOD1 mice. We and others had previously shown mitochondrial dysfunction accompanied by ATP depletion and UCP3 upregulation in skeletal muscle of mSOD1 mice (16) and ALS patients (28-31). It is thus probable that this mitochondrial impairment could underlie the increased energy needs of skeletal muscle as reflected by the increased rates of glucose uptake (6) and TG clearance shown here.

The mechanisms recruited by mSOD1 muscles to increase lipid uptake are not completely understood. We observed the elevated expression of enzymes and transporters involved in lipid metabolism, including VLDLR, LDLR and FAT/CD36, but failed to demonstrate an increase in LPL activity in mSOD1 muscles. In contrast, muscle LPL mRNA levels, while unchanged in presymptomatic G86R mice, appeared increased at disease onset as well as in both presymptomatic and diseased G93A mice, suggesting that LPL might be involved in promoting a higher lipid uptake. These results contrasted with the transcriptional downregulation of LPL in denervated muscles, which suggests that other mechanisms distinct from pure denervation are influencing muscle pathology in the transgenic mice.

There are several potential explanations for the lack of an increase in LPL activity in mSOD1 muscles. First, we assayed total LPL activity, while only a fraction of it, *i.e.* the heparin-releasable fraction, is biologically active. Thus, measurement of total activity could have masked the action of the active LPL. Second, LPL activity *in vivo* is regulated by numerous factors, in particular plasma proteins, such as apoCII (activator) or apoCIII (inhibitor), that are are lost in our assay. Third, genes known to increase TG-rich lipoprotein clearance through LPL, such as VLDLR (32-34) or FAT/CD36 (35, 36), were upregulated in presymptomatic mSOD1 mice muscles. The increased expression of these genes is per se sufficient to infer increased lipid uptake in muscle and, notably, FAT/CD36 deficiency was recently reported to decrease TG-rich lipoprotein clearance without affecting post-heparin LPL activity (35), consistent with a role of CD36 in increasing local LPL activity without increasing either expression or assayable activity of LPL. Altogether, our data suggest that LPL activity could actually be increased in mutant SOD1 mouse muscles through yet unknown local factors, which may include CD36. In all, skeletal muscle hypermetabolism is likely to trigger the observed aberrant decreased postprandial lipidemia in mSOD1 mice. In this scenario, the protective potential of high fat diet might be interpreted as an increased supply of high energy nutrients to the muscle, thus compensating its dysfunction.

A recent report claimed the lack of involvement of skeletal muscle in mSOD1 triggered pathology (37). These authors used floxed mSOD1 mice (15) and ablated mSOD1 expression in skeletal muscles using targeted CRE expression. However, recombination occurred significantly only in one of the two muscles tested. Most importantly, the efficiency of recombination in respiratory muscles was not provided, although it is thought that their failure trigger the death of the mice. Last, neither motor neuron counting nor neuromuscular junction morphology were provided, thus weaknessing the conclusions drawn from these experiments. In all, the question of the contribution of skeletal muscle mSOD1 expression remains open. It is however clear that skeletal muscle hypermetabolism might also be triggered by mutant SOD1 expression in other cells. Further research is needed to elucidate this point.

Our report gives clues for a nutritional management of ALS patients, suggesting that increasing calorie intake might increase survival and that hypolipidemic drugs such as fibrates, cholestyramine or statins should be avoided in these patients, since decreasing lipidemia is likely to exacerbate the ALS condition. Contrary to animal models of Parkinson's disease (38) or Huntington's disease (39), where caloric restriction has been shown to alleviate symptoms, caloric restriction shortens disease duration in mSOD1 mice (21). Our studies on mSOD1 mice suggest that maintaining body mass index should slow disease progression.

**Acknowlegdements**

We thank the expert technical assistance of Marie José Ruivo, Annie Picchinenna, Nicole Guivier and David Chadeyron. We also acknowledge Dr Christian KOEHL (Laboratoire de Biochimie Générale et Spécialisée, Hôpitaux universitaires de Strasbourg) and Dr Andoni ECHANIZ-LAGUNA for help with triglycerides and cholesterol dosages. The study was supported by grants from Association pour l'Etude de la Culture d'Embryons et des Thérapeutiques des Maladies du Système Nerveux (ACE), Région Alsace and Fondation pour la Recherche Médicale (FRM) to A.F.; Fondation pour la Recherche sur le Cerveau (FRC) to L.D.; Association pour la Recherche sur la Sclérose Latérale Amyotrophique (ARS) to J.L.G.D.A.; and Association Française contre les Myopathies (AFM), Alsace Biovalley, ARS and Association pour la Recherche et le Développement de Moyens de Lutte contre les Maladies Neurodégénératives (AREMANE) to J.P.L.

# References

1.  Pedersen, W.A., and E.R. Flynn. 2004. Insulin resistance contributes to aberrant stress responses in the Tg2576 mouse model of Alzheimer's disease. *Neurobiol Dis* **17**:500-506.

2.  Pedersen, W.A., P.J. McMillan, J.J. Kulstad, J.B Leverenz, S. Craft, and G.R. Haynatzki. 2006. Rosiglitazone attenuates learning and memory deficits in Tg2576 Alzheimer mice. *Exp Neurol.* **199**:265-273

3.  Bjorkqvist, M., M. Fex, E. Renstrom, N. Wierup, A. Petersen, J. Gil, K. Bacos, N. Popovic, J.Y. Li, F. Sundler, P. Brundin, and H. Mulder. 2005. The R6/2 transgenic mouse model of Huntington's disease develops diabetes due to deficient beta-cell mass and exocytosis. *Hum Mol Genet* **14**:565-574.

4.  Fain, J.N., N.A. Del Mar, C.A. Meade, A. Reiner, and D. Goldowitz. 2001. Abnormalities in the functioning of adipocytes from R6/2 mice that are transgenic for the Huntington's disease mutation. *Hum Mol Genet* **10**:145-152.

5.  Hurlbert, M.S., W. Zhou, C. Wasmeier, F.G. Kaddis, J.C. Hutton, and C.R. Freed. 1999. Mice transgenic for an expanded CAG repeat in the Huntington's disease gene develop diabetes. *Diabetes* **48**:649-651.

6.  Dupuis, L., H. Oudart, F. Rene, J.L. Gonzalez de Aguilar, and J.P. Loeffler. 2004. Evidence for defective energy homeostasis in amyotrophic lateral sclerosis: benefit of a high-energy diet in a transgenic mouse model. *Proc Natl Acad Sci U S A* **101**:11159-11164.

7.  Weydt, P., V.V. Pineda, A.E. Torrence, R.T. Libby, T.F. Satterfield, E.R. Lazarowski,

M.L. Gilbert, G.J. Morton, T.K. Bammler, A.D. Strand, L. Cui, R.P. Beyer, C.N. Easley, A.C. Smith, D. Krainc, S. Luquet, I.R. Sweet, M.W. Schwartz, and A.R. La Spada. 2006. Thermoregulatory and metabolic defects in Huntington's disease transgenic mice implicate PGC-1alpha in Huntington's disease neurodegeneration. *Cell Metab* **4**:349-362.

8. Pasinelli, P., and R.H. Brown. 2006. Molecular biology of amyotrophic lateral sclerosis: insights from genetics. *Nat Rev Neurosci* **7**:710-723.

9. Gonzalez de Aguilar, J.L., A. Echaniz-Laguna, A. Fergani, F. René, V. Meininger, J.P. Loeffler, and L. Dupuis. 2006. Amyotrophic lateral sclerosis: all roads lead to Rome. *J Neurochem* in press.

10. Ripps, M.E., G.W. Huntley, P.R. Hof, J.H. Morrison, and J.W. Gordon. 1995. Transgenic mice expressing an altered murine superoxide dismutase gene provide an animal model of amyotrophic lateral sclerosis. *Proc Natl Acad Sci U S A* **92**:689-693.

11. Gurney, M.E., H. Pu, A.Y. Chiu, M.C. Dal Canto, C.Y. Polchow, D.D. Alexander, J. Caliendo, A. Hentati, Y.W. Kwon, and H.X. Deng. 1994. Motor neuron degeneration in mice that express a human Cu,Zn superoxide dismutase mutation. *Science* **264**:1772-1775.

12. Wong, P.C., C.A. Pardo, D.R. Borchelt, M.K. Lee, N.G. Copeland, N.A. Jenkins, S.S. Sisodia, D.W. Cleveland, and D.L. Price. 1995. An adverse property of a familial ALS-linked SOD1 mutation causes motor neuron disease characterized by vacuolar degeneration of mitochondria. *Neuron* **14**:1105-1116.

13. Rosen, D.R., T. Siddique, D. Patterson, D.A. Figlewicz, P. Sapp, A. Hentati, D. Donaldson, J. Goto, J.P. O'Regan, H.X. Deng, Z. Rahmani, A. Krizus, D. McKenna-

Yasek, A. Cayabyab, S.M. Gaston, R. Berger, R.E. Tanzi, J.J. Halperin, B. Herzfeldt, R. Van den Bergh, W.Y. Hung, T. Bird, G. Deng, D.W. Mulder, C. Smyth, G.N. Laing, E. Soriano, M.A. Pericak–Vance, J. Haines, G.A. Rouleau, J.S. Gusella, H.R. Horvitz, and R.H. Brown. 1993. Mutations in Cu/Zn superoxide dismutase gene are associated with familial amyotrophic lateral sclerosis. *Nature* **362**:59-62.

14. Clement, A.M., M.D. Nguyen, E.A. Roberts, M.L. Garcia, S. Boillee, M. Rule, A.P. McMahon, W. Doucette, D. Siwek, R.J. Ferrante, R.H. Brown, J.P. Julien, L.S. Goldstein, and D.W. Cleveland. 2003. Wild-type nonneuronal cells extend survival of SOD1 mutant motor neurons in ALS mice. *Science* **302**:113-117.

15. Boillee, S., K. Yamanaka, C.S. Lobsiger, N.G. Copeland, N.A. Jenkins, G. Kassiotis, G. Kollias, and D.W. Cleveland. 2006. Onset and progression in inherited ALS determined by motor neurons and microglia. *Science* **312**:1389-1392.

16. Dupuis, L., F. di Scala, F. Rene, M. de Tapia, H. Oudart, P.F. Pradat, V. Meininger, and J.P. Loeffler. 2003. Up-regulation of mitochondrial uncoupling protein 3 reveals an early muscular metabolic defect in amyotrophic lateral sclerosis. *Faseb J* **17**:2091-2093.

17. Desport, J.C., P.M. Preux, L. Magy, Y. Boirie, J.M. Vallat, B. Beaufrere, and P. Couratier. 2001. Factors correlated with hypermetabolism in patients with amyotrophic lateral sclerosis. *Am J Clin Nutr* **74**:328-334.

18. Kasarskis, E.J., S. Berryman, J.G. Vanderleest, A.R. Schneider, and C.J. McClain. 1996. Nutritional status of patients with amyotrophic lateral sclerosis: relation to the proximity of death. *Am J Clin Nutr* **63**:130-137.

19. Gonzalez de Aguilar, J.L., L. Dupuis, H. Oudart, and J.P. Loeffler. 2005. The metabolic hypothesis in amyotrophic lateral sclerosis: insights from mutant Cu/Zn-superoxide dismutase mice. *Biomed Pharmacother* **59**:190-196.

20. Mattson, M.P., R.G. Cutler, and S. Camandola. 2007. Energy intake and amyotrophic lateral sclerosis. *Neuromolecular Med* **9**:17-20.

21. Pedersen, W.A., and M.P. Mattson, 1999. No benefit of dietary restriction on disease onset or progression in amyotrophic lateral sclerosis Cu/Zn-superoxide dismutase mutant mice. *Brain Res* **833**:117-120.

22. Habold, C., C. Foltzer-Jourdainne, Y. Le Maho, J.H. Lignot, and H. Oudart, 2005. Intestinal gluconeogenesis and glucose transport according to body fuel availability in rats. *J Physiol.* **566**:575-86.

23. Hocquette, J.F., B. Graulet, and T. Olivecrona, 1998. Lipoprotein lipase activity and mRNA levels in bovine tissues. *Comp Biochem Physiol B Biochem Mol Biol* **121**:201-212.

24. Toepfer, M., C. Folwaczny, A. Klauser, R.L. Riepl, W. Muller-Felber, and D. Pongratz, 1999. Gastrointestinal dysfunction in amyotrophic lateral sclerosis. *Amyotroph Lateral Scler Other Motor Neuron Disord* **1**:15-19.

25. Eberle, D., B. Hegarty, P. Bossard, P. Ferre, and F. Foufelle. 2004. SREBP transcription factors: master regulators of lipid homeostasis. *Biochimie* **86**:839-848.

26. Russell, D.W. 2003. The enzymes, regulation, and genetics of bile acid synthesis. *Annu Rev Biochem* **72**:137-174.

27. Desport, J.C., F. Torny, M. Lacoste, P.M. Preux, and P. Couratier. 2005. Hypermetabolism in ALS: correlations with clinical and paraclinical parameters. *Neurodegener Dis.* **2**:202-207.

28. Echaniz-Laguna, A., J. Zoll, E. Ponsot, B. N'Guessan, C.Tranchant, J.P. Loeffler, and

E. Lampert. 2006. Muscular mitochondrial function in amyotrophic lateral sclerosis is progressively altered as the disease develops: a temporal study in man. *Exp Neurol* **198**:25-30.

29. Vielhaber, S., K. Winkler, E. Kirches, D. Kunz, M. Buchner, H. Feistner, C.E. Elger, A.C. Ludolph, M.W. Riepe, and W.S. Kunz. 1999. Visualization of defective mitochondrial function in skeletal muscle fibers of patients with sporadic amyotrophic lateral sclerosis. *J Neurol Sci* **169**:133-139.

30. Vielhaber, S., D. Kunz, K. Winkler, F.R. Wiedemann, E. Kirches, H. Feistner, H.J. Heinze, C.E. Elger, W. Schubert, and W.S. Kunz. 2000. Mitochondrial DNA abnormalities in skeletal muscle of patients with sporadic amyotrophic lateral sclerosis. *Brain* **123**:1339-1348.

31. Krasnianski, A., M. Deschauer, S. Neudecker, F.N. Gellerich, T. Muller, B.G. Schoser, M. Krasnianski, and S. Zierz. 2005. Mitochondrial changes in skeletal muscle in amyotrophic lateral sclerosis and other neurogenic atrophies. *Brain* **128**:1870-1876.

32. Espirito Santo, S.M., P.C. Rensen, J.R. Goudriaan, A. Bensadoun, N. Bovenschen, P.J. Voshol, L.M. Havekes, and B.J. van Vlijmen. 2005. Triglyceride-rich lipoprotein metabolism in unique VLDL receptor, LDL receptor, and LRP triple-deficient mice. *J Lipid Res* **46**:1097-1102.

33. Goudriaan, J.R., S.M. Espirito Santo, P.J. Voshol, B. Teusink, K.W. van Dijk, B.J. van Vlijmen, J.A. Romijn, L.M. Havekes, and P.C. Rensen. 2004. The VLDL receptor plays a major role in chylomicron metabolism by enhancing LPL-mediated triglyceride hydrolysis. *J Lipid Res* **45**:1475-1481.

34. Yagyu, H., E.P. Lutz, Y. Kako, S. Marks, Y. Hu, S.Y. Choi, A. Bensadoun, and I.J. Goldberg. 2002. Very low density lipoprotein (VLDL) receptor-deficient mice have reduced lipoprotein lipase activity. Possible causes of hypertriglyceridemia and reduced body mass with VLDL receptor deficiency. *J Biol Chem* **277**:10037-10043.

35. Drover, V.A., M. Ajmal, F. Nassir, N.O. Davidson, A.M. Nauli, D. Sahoo, P. Tso, and N.A. Abumrad. 2005. CD36 deficiency impairs intestinal lipid secretion and clearance of chylomicrons from the blood. *J Clin Invest* **115**:1290-1297.

36. Ibrahimi, A., A. Bonen, W.D. Blinn, T. Hajri, X. Li, K. Zhong, R. Cameron, and N.A. Abumrad. 1999. Muscle-specific overexpression of FAT/CD36 enhances fatty acid oxidation by contracting muscle, reduces plasma triglycerides and fatty acids, and increases plasma glucose and insulin. *J Biol Chem* **274**:26761-26766.

37. Miller, T.M., S.H. Kim, K. Yamanaka, M. Hester, P. Umapathi, H. Arnson, L. Rizo, J.R. Mendell, F.H. Gage, D.W. Cleveland, and B.K. Kaspar. 2006. Gene transfer demonstrates that muscle is not a primary target for non-cell autonomous toxicity in familial amyotrophic lateral sclerosis. *Proc Natl Acad Sci U S A* **103**:19546-19551.

38. Maswood, N., J. Young, E. Tilmont, Z. Zhang, D.M. Gash, G.A. Gerhardt, R. Grondin, G.S. Roth, J. Mattison, M.A. Lane, R.E. Carson, R.M. Cohen, P.R. Mouton, C. Quigley, M.P. Mattson, and D.K. Ingram. 2004. Caloric restriction increases neurotrophic factor levels and attenuates neurochemical and behavioral deficits in a primate model of Parkinson's disease. *Proc Natl Acad Sci U S A* **101**:18171-18176.

39. Duan, W., Z. Guo, H. Jiang, M. Ware, X.J. Li, and M.P. Mattson. 2003. Dietary restriction normalizes glucose metabolism and BDNF levels, slows disease progression, and increases survival in huntingtin mutant mice. *Proc Natl Acad Sci U S*

*A* **100**:2911-2916.

**Figure legends :**

**Figure 1 : Decreased postprandial lipidemia in mSOD1 mice**

A-C :  Plasma TG (A), cholesterol (B) and total lipids (C) in fed, fasted or olive oil-gavaged wild-type (empty columns) and mSOD1 (black columns) mice. n = 10-15 mice for panels A and B ; n = 6 mice for panel C. *, $p < 0.05$ *versus* wild-type.

D:    Plasma cholesterol in wild-type and mSOD1 mice fed with either chow diet or high fat diet (HFD). n = 10 mice *, $p < 0.05$ *versus* corresponding wild-type.

**Figure 2: FPLC fractionation of lipoproteins in mSOD1 mice**

A :   Representative FPLC profiles of fed wild-type (grey line) and mSOD1 (black line) mice. Note the depletion of VLDL and LDL cholesterol fractions in mSOD1 mice.

B:    Quantification of the experiments presented in D. n = 6 mice *, $p < 0.05$ *versus* wild-type. Column legends as in figure 1.

C: Quantification of FPLC profiles of 4h-fasted wild-type (grey line) and mSOD1 (black line) mice. n = 3 pools of 3 mice. Column legends as in figure 1.

**Figure 3: Increased intestinal absorption in mSOD1 mice**

A :   Real-time RT-PCR analysis of SOD1 expression in wild-type (Wt, empty columns), non-symptomatic mSOD1 (NS, black columns) and symptomatic mSOD1 (onset, OS, black columns). n= 5 mice. *, $p < 0.05$ *versus* wild-type. mRNA levels are expressed in arbitrary units (A.U.) and are normalized to 18S rRNA.

B: Gastric emptying of methylene blue in non-symptomatic mSOD1 (black columns) and wild-type (empty columns) mice. Animals were sacrificed 30 minutes after gavage, and the remaining methylene blue in the stomach was assayed spectrophotometrically and compared to reference mice sacrificed 1 minute after gavage. n = 9 mice. No difference is observed between the two groups.

C: Intestinal transit of methylen blue in the same mice presented in A. No difference is observed between the two groups.

D: Scheme depicting the experimental paradigm used in E and F. Triglycerides (TG) are perfused into the intestinal segment, hydrolyzed by intestinal enzymes into non-esterified fatty acids (NEFA), absorbed in enterocytes and re-converted in TG before chylomicron secretion.

E-F: Levels of TG and NEFA in the intestinal lumen as a function of time of perfusion. Note that TG decrease faster in mSOD1 mice (black line) than in wild-type mice (grey line) (panel E), whereas steady state levels of NEFA remained constantly lower in mSOD1 mice but increased gradually in wild-type mice (panel F). n = 6-8 mice. *, $p < 0.05$ versus corresponding perfusion time in wild-type mice.

G : Glucose absorption rate in wild-type (empty column) and non-symptomatic mSOD1 mice (black column). No significant difference is noted between the groups. n = 3-4 mice.

**Figure 4 : Hepatic metabolism in mSOD1 mice**

A :  Real-time RT-PCR analysis of the indicated genes in the liver of wild-type (Wt, empty columns), non-symptomatic (NS) or symptomatic at onset (OS) mSOD1 mice (black columns). n = 7 mice. *, $p < 0.05$ *versus* wild-type mice. mRNA levels are expressed in arbitrary units (A.U.) and are normalized to 18S rRNA.

B : Representative Oil red O stainings of liver sections from wild-type (Wt), non-symptomatic (NS mSOD1) and symptomatic at onset (OS mSOD1) mSOD1 mice. n = 3-5 mice.

**Figure 5: lipoprotein production and clearance in mSOD1 mice**

A: Determination of vLDL-TG production rates in wild-type (Wt, grey line) and non-symptomatic mSOD1 mice (mSOD1, black line) After a 4-hour fast, tyloxapol was injected iv and plasma samples were assayed for TG content at the indicated time points. Note that TG increases are indistinguishable between the two groups of animals, thus showing that vLDL secretion is normal in mSOD1 animals. n = 6 mice.

A : Plasma TG content during a fat-loading test in wild-type (Wt, grey line) and non-symptomatic mSOD1 mice (mSOD1, black line) (n = 6 mice). After a 4-hour fast, mice were gavaged with 400 μl olive oil. Before and after fat loading, blood was collected serially, and plasma TG levels were measured. The experiments was repeated three times with independent cohorts of mice and yielded similar results. *$P < 0.05$ *versus* control littermates.

**Figure 6 : expression of genes involved in lipid uptake in skeletal muscles of mSOD1 mice**

A : Real-time RT-PCR analysis of LPL expression in the hindlimb skeletal muscles of wild-type (Wt, empty columns), non-symptomatic (NS) or symptomatic at onset (OS) mSOD1mice (black columns). mRNA levels of LPL were also measured in axotomized mice (Axo) at the ipsilateral (I) and contralateral (C) side of the lesion. Sham-operated mice served as control. *, $p < 0.05$ *versus* corresponding wild-type mice at the same time point. mRNA levels are expressed in arbitrary units (A.U.) and are normalized to 18S rRNA.

B : Total LPL activity in hindlimb muscles of wild-type (Wt, empty column) and non-symptomatic mSOD1 mice (black column). n = 8 mice. No significant difference was noted between the two groups.

C : Real-time RT-PCR analysis of the indicated genes in the hindlimb skeletal muscles of wild-type (Wt, empty columns), non-symptomatic (NS) or symptomatic at onset (OS) mSOD1mice (black columns). *, p < 0.05 *versus* corresponding wild-type mice. mRNA levels are expressed in arbitrary units (A.U.) and are normalized to 18S rRNA.

D: Real-time RT-PCR analysis of the indicated genes in the hindlimb skeletal muscles of wild-type (Wt, empty columns), non-symptomatic (NS) and symptomatic at onset (OS) G93A mice (black columns). n = 6 mice. *, p < 0.05 *versus* corresponding wild-type mice. mRNA levels are expressed in arbitrary units (A.U.) and are normalized to 18S rRNA.

Fergani and collaborators, figure 1

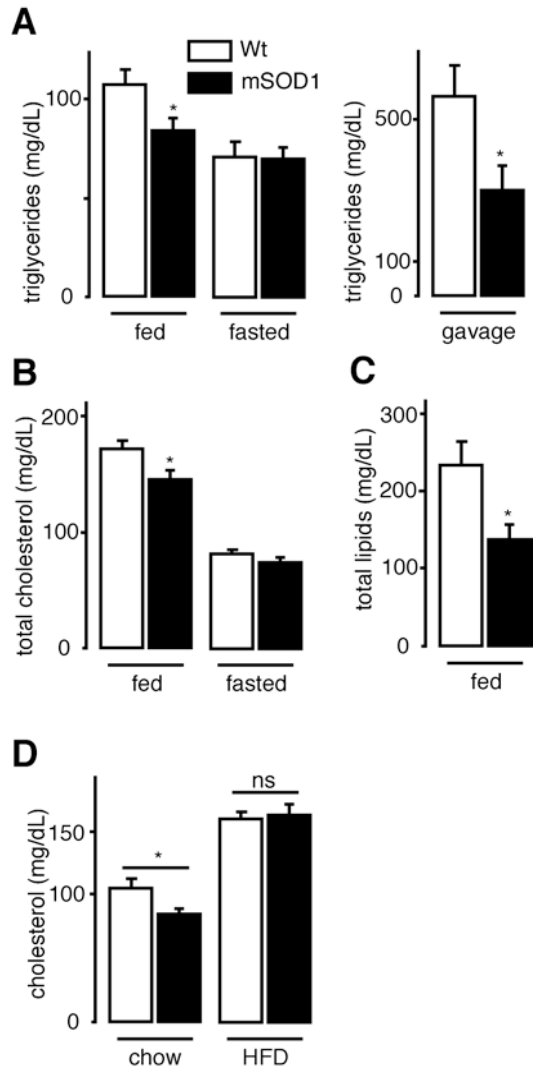

**Fergani and collaborators, figure 2**

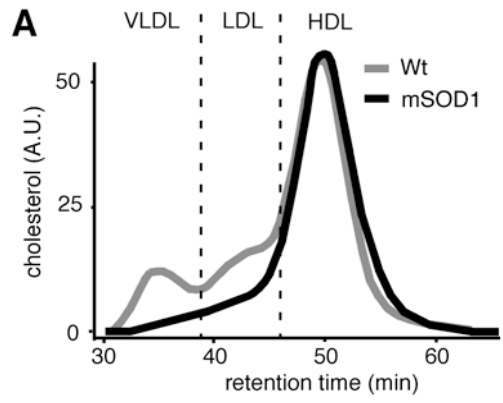

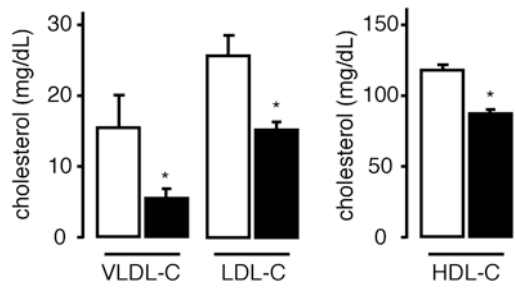

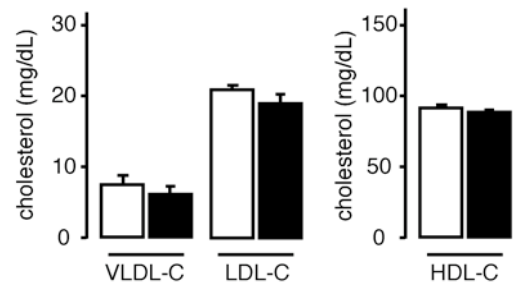

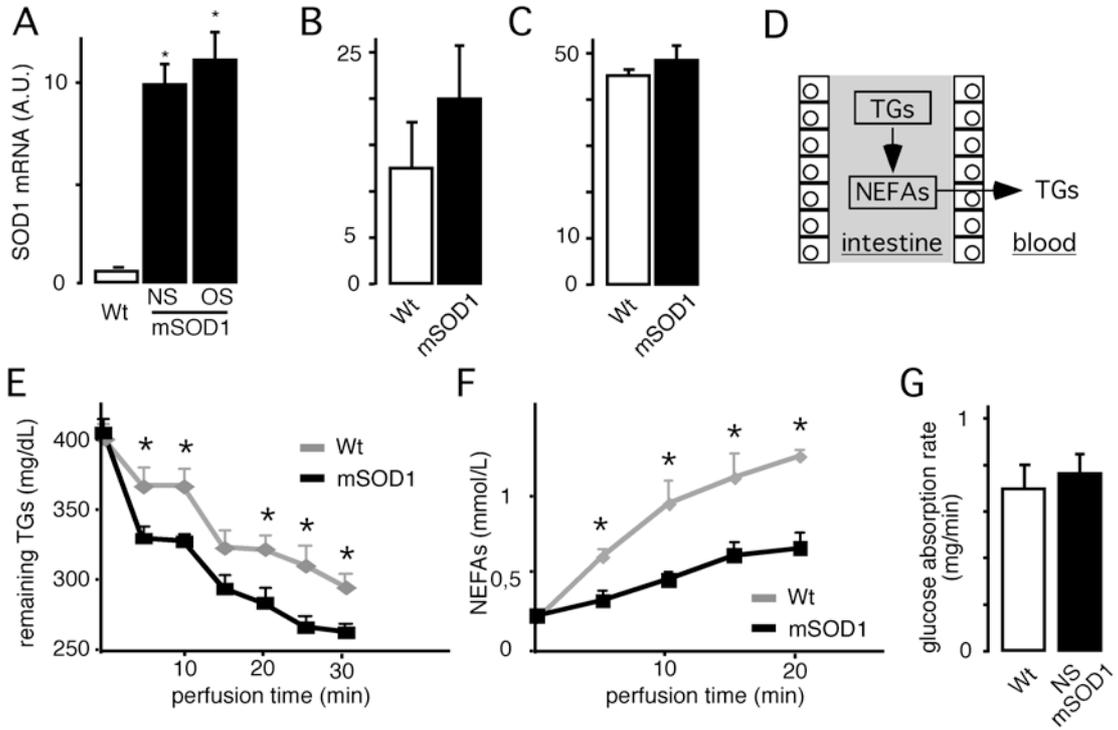

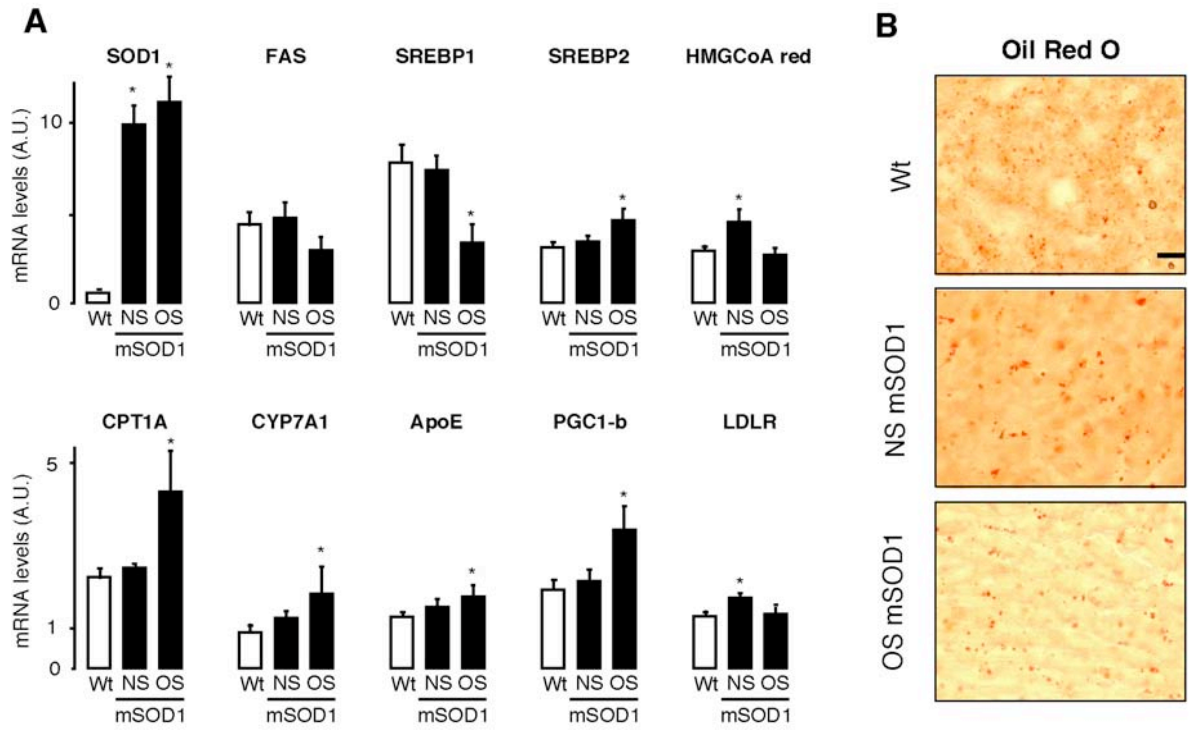

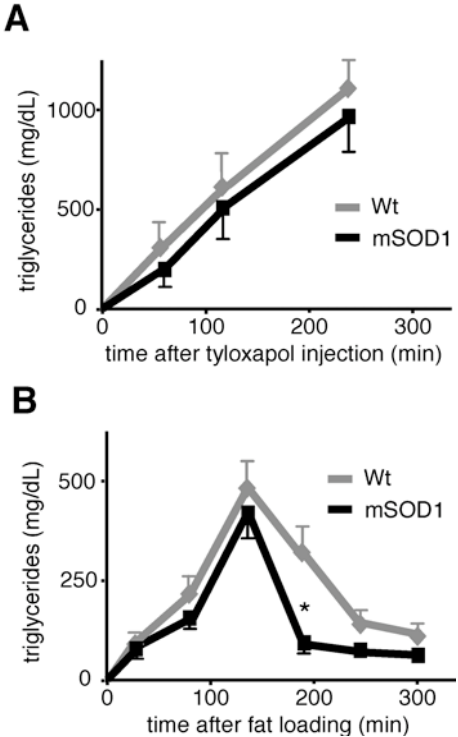

Fergani and collaborators, figure 5

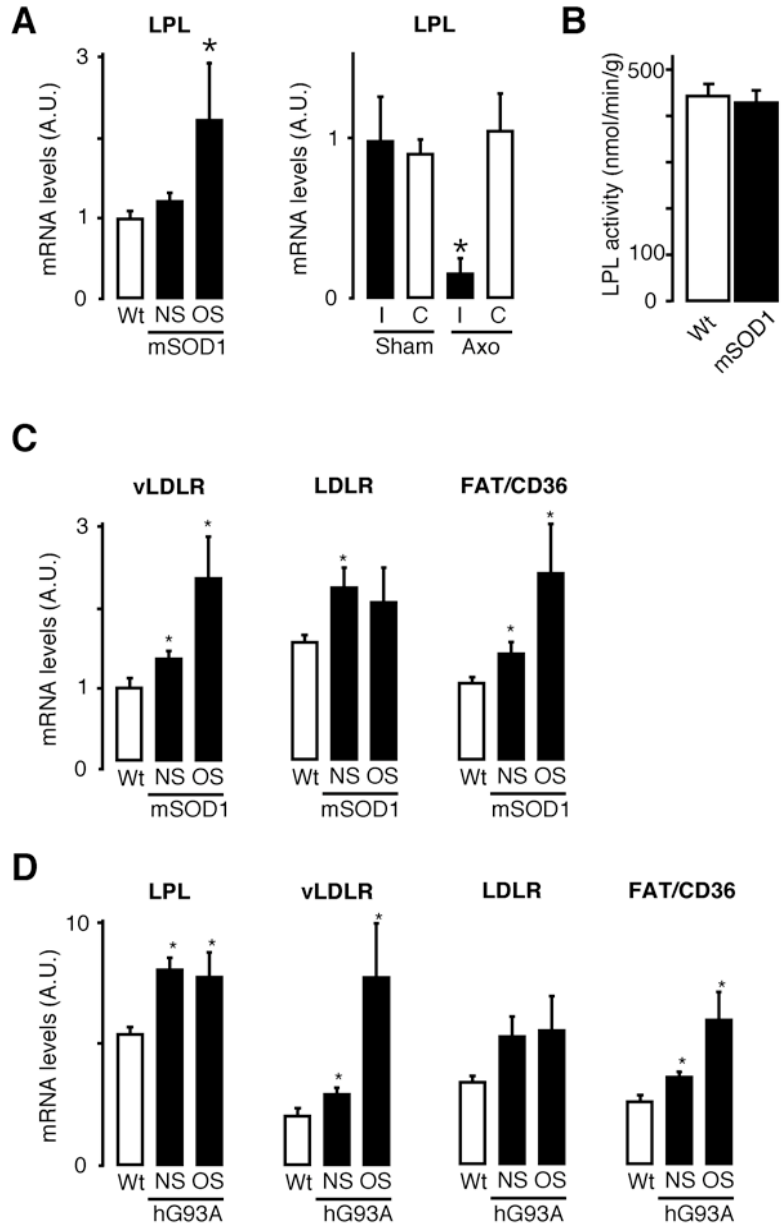

Fergani and collaborators, figure 6